# Robust internal model control approach for position control of systems with sandwiched backlash

Yoav Vered[a] and Stephen J. Elliott[b]
Institute of Sound and Vibration Research, University of Southampton, SO17 1BJ, United Kingdom

*This paper describes the design of a robust controller for position control in systems with sandwiched backlash. The backlash, which is nonsmooth and nonlinear, is inevitable in the operation of many systems, but it can have destructive effects on the stability and performance of feedback systems. In this work, a robust controller is designed using a modified linear internal model control framework. Different controller architectures are considered and compared based on an experimental case study. The experimental testbased is composed of a three-platform structure driven by a stepper motor. The backlash is introduced into the system in a non-destructive and controllable manner by closing an internal nonlinear feedback loop around the stepper motor. The robustness of the designed controller to a large amount of backlash is verified experimentally, and while the stability is maintained, some residual vibrations are observed. The effects on the residual vibration levels of including nonlinear elements in the controller and changing the controller's settling time are also examined experimentally. The robustness to changes and mismodelling of the linear system, with and without the backlash, is described, as is the tracking of a smooth sinusoidal command signal with a growing amount of backlash. Based on the case study, it is concluded that combining the linear internal model control design method with a small dead zone results in a highly robust controller both with respect to the backlash and to changes in the linear system, which ensure stability and good performance. The required robustness is achieved by tuning the controller's settling time and the dead zone width parameters.*

---

[a] Corresponding Author, electronic mail: V.Yoav@soton.ac.uk
[b] Electronic mail: S.J.Elliott@soton.ac.uk

# 1. Introduction

Feedback in position control is crucial to overcome the effect of disturbances and to ensure convergence and robustness under structural uncertainties. When considering linear plants, known robust control methods are available to design the controller. However, position control mechanisms that use a geared motor are generally not linear [1]. Backlash, a nonsmooth and nonlinear operator, is inevitable for their proper operation and to enable the transmission of displacement and mechanical power. The backlash may also grow over time, due to the wear and tear of transmission parts. The growing backlash can deteriorate the performance and cause stability loss in the form of limit cycle oscillations [2]. Most real-life position control with backlash applications use a model-free controller, such as a PID controller [3]. It is, therefore, attractive to investigate the design of a simple-structured robust model-based controller for such systems.

Backlash is inevitable in the operation of any mechanical transmission and is modelled frequently by a clearance between the different transmission's sides [4,5]. In practice, there is generally a dynamic system on each side of the transmission, which might be linear or nonlinear. Here, the case where the backlash clearance is situated between two linear systems will be considered. Moreover, it is assumed that the backlash does not affect the input but only the output. This case is referred to herein as a sandwiched backlash [3]. Often, passive solutions are used to minimise the effects of the backlash, such as the anti-backlash nut, which aims to reduce the clearance with a preloaded spring. However, though the amount of backlash might be negligible initially, it can grow over time [6]. When the backlash grows, the feedback system can lose its stability, especially if controlled by a PID controller [7]. Other methods to control systems with backlash were also previously considered [4]. But most effective control methods result in a complex structured controller; examples are given in [3,4,8–10].

An observer-based feedback controller is one of the standard model-based control schematics. The idea is to use a reduced-order model of the system and possibly the input control signal and output measurements as part of the feedback loop. The combination of the model and measurements is usually referred to as an observer. The observer's goal can be to estimate the internal states [11,12], to reduce the effect of disturbance or measurement noise, to estimate external disturbance signals [13], to estimate the modelling error [14], or a combination of them all. Optimal model-based robust feedback control design is a well-developed field, with popular optimal and robust design methods for linear systems such as the $H_2$, $H_\infty$, and $H_\infty$ loop-shaping [15]. However, these methods are not entirely transparent, meaning the physical interpretation can be lost. Instead, the designer aims to minimise some cost function, which might affect the actual system, but will not necessarily have the desired effect.

The internal model control (IMC) architecture [16,17] and disturbance observer-based control (DOB) [13] are popular methods due to their simplicity, flexibility, and transparency. The main idea in both paradigms is to introduce an estimator in the feedback loop, which reflect either the amount of model uncertainty (IMC) or disturbance (DOB). The idea is similar to the observer-based state feedback control. However, in these methods, the observer is not designed such that the estimation error is asymptotically stable (meaning it converges to zero) as in the state estimator or Kalman filter. Instead, the estimation error is used to drive the feedback control and to ensure that the output follows the desired input command signal. In [18], an overview of the similarities and differences between the IMC and DOB is given. The main advantage of the DOB compared to the IMC is its digital implementation. However, the filter design procedure of the IMC transforms the feedback design problem into a feedforward one, for which elegant solutions exist.

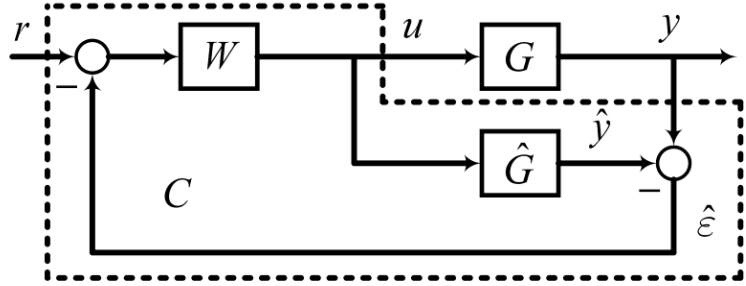
Figure 1: IMC-based feedback position controller.

The IMC concept, shown by the block diagram in Figure 1, is well-established for the design of feedback controllers for internally stable systems [14,19], and it was introduced by both Zames [19] and Garcia and Morari [20]. In the block diagram $r, u,$ and $y$ are the command (reference), control, and output signals, $G$ the physical plant response and $\hat{G}$ its internal model. $\hat{y}$ and $\hat{\varepsilon}$ are the controller internal output estimation and the estimation error signals. $W$ is the controller prefilter, and the equivalent SISO controller $C_e$, from the tracking error $e = r - y$ to the control signal, $u$, is equal to $W/(1 - W\hat{G})$. The advantage of using an IMC for reference tracking or position control problems is in the fact that when the plant model is accurate, so that $\hat{G} = G$, an open-loop feedforward filter can be designed, which generally is a more straightforward task than designing a feedback controller. However, unlike the feedforward control, the IMC can then be used to compensate for changes in the plant response and attenuate the effect of input and output disturbances. Also, based on the small-gain theorem, it is possible to design a filter that ensures the stability of the closed loop as long as the modelling error can be bounded [14].

Algorithms to design the IMC controller have been suggested both for minimum and nonminimum phase plants [21]. The method presented in [21] aims to cancel the plant's poles, minimum-phase zeros, and the projection of the nonminimum-phase zeros on the left-hand side complex plane while ensuring robustness by introducing a low-pass filter. This is equivalent to the feedforward model reference design method. However, doing so can result in significantly high control power, especially when cancelling higher system modes. The required control power can be reduced by truncating the original plant to include only the dominant modes [22]. Furthermore, since the problem is cast as a feedforward problem, an input-shaping approach [23,24] could also be used, ensuring robustness and bounds on the settling time but losing some of the physical interpretation transparency.

As has been discussed in the literature [15] (Theorem 12.7) for a stable linear plant, $G,$ the parametrization of all stabilising controllers via the Youla–Kučera parametrisation results in an IMC structured controller. For unstable plants, the parametrization becomes more cumbersome but can be obtained using coprime factorisation [25,26]. However, it can be more straightforward to use a cascade controller instead by first stabilising the unstable plant using a simple structured internal controller and then applying the IMC design to the stable plant as the outer controller. This can be especially advantageous for MIMO systems when only some subsystems are unstable. An example is a servomechanism driven by an electrical motor, similar to the one used and described in this paper.

Nonlinear forms of IMC [27] have also been discussed in the literature. Also, adaptive controllers have been considered, where the internal model's parameters and the prefilter are estimated simultaneously in real-time [28,29]. Datta's monograph [30] gives an extensive summary of both stationary-based and adaptive-based nonlinear IMC design methodologies. However, such approaches often result in a complex structure for the controller, whose implementation is not straightforward, so this approach is not followed here.

The focus of this paper is on designing a linear IMC for an estimated linear system model and then adding the required nonlinear elements to the controller, similar to the approach of [22,31]. In [22], a nonlinear backlash estimator was added to the inner angle-control feedback loop and to the internal model. While in [32], in addition to the use of the backlash estimator in the internal model, an approximated backlash inverse model is also added to the prefilter. Both of the proposed controllers of [22,31] were evaluated numerically and showed considerable improvement over the linear IMC. However, the need to implement the inverse backlash model can result in a large control signal amplitude and a complex controller structure.

The novelty of the current work is the design of an IMC-based controller for the sandwiched backlash that is assessed and evaluated based on an extensive experimental case study. A three-platform structure was used as the experimental testbed. The relative motion between the bottom and middle platform was controlled using an accurate stepper motor. To enable the non-destructive evaluation for different levels of backlash, an additional nonlinear feedback loop was closed around the stepper motor. Thus, virtual dynamics and backlash are introduced into the experimental physical system to allow a user-specified amount of nonlinearity to be introduced.

The aim of the proposed controller design framework is to have a meaningful and straightforward physical controller architecture. The proposed architectures and design parameters are assessed and compared using the experimental testbed. Based on the results, it is found that the robustness for linear structural uncertainties is crucial for the robustness to backlash. The former can be obtained to the desired level by tuning the controller's settling time at the design step. Alternatively, this task can be done automatically by employing a digital twin [7], for example.

The paper is organized as follows: The experimental testbed description, its analytical model derivation, and the experimental identification of its model parameters are described in Section 2. The analytical backlash model, the virtual backlash nonlinear control and experimental verification are shown in Section 3. The linear IMC controller is described and analysed analytically and experimentally in Section 4. Then the addition of nonlinear components is examined experimentally in Section 5. Finally, Sections 6 and 7 report on experimental case studies carried out to investigate the proposed controller robustness to changes in the nominal plant and the effect of the designed settling time.

## 2. Experimental testbed

An experimental three-platform structure is used to verify and check the robustness of the different controllers' schematics described in the following sections. Figure 2(a) and Figure 2(b) present a photograph of the experimental system used as the testbed and of the study case flow diagram. The system comprises three stainless-steel plates connected to each other and the ground by four aluminium beams. To increase the beam's damping, a viscoelastic damping material was added on top of each beam. The system is driven by a Ding double-stacked 24V hybrid stepper motor with a 0.1" (2.54mm) pitch per revolution located on top of the first plate and capable of driving the relative degree of freedom between the bottom and middle platforms. A 2500 pulse-per-revolution encoder is used to sense the stepper motor's angle, and a Keyence LK-G32 laser triangulation sensor is used to measure the position of the top platform. A Leadshine DM805-AI driver was used for the stepper motor in the pulse direction configuration. A DS1103 PPC dSPACE controller board was used to measure the sensors, generate the excitation signal, and realise the controller and virtual dynamics that will be introduced later in the paper. The sampling frequency was $F_s = 10^4$ Hz corresponds to a sample time of 0.1 milliseconds. A 15-bit analogue to digital converter was used for the Keyence signal, digital outputs were used to feed the stepper motor's driver, and the stepper angle was measured using the incremental encoder interface connector.

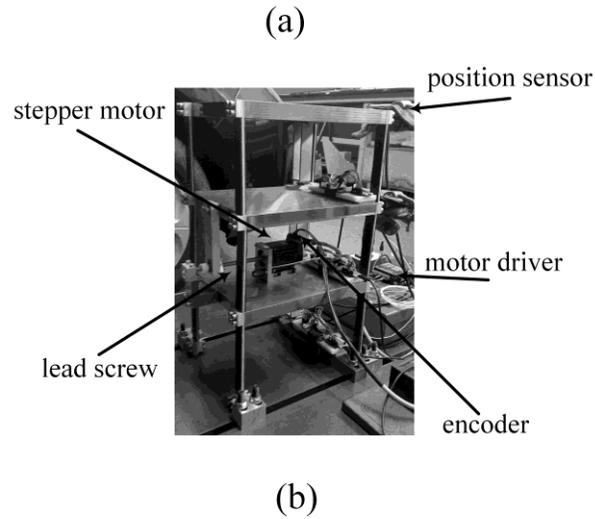

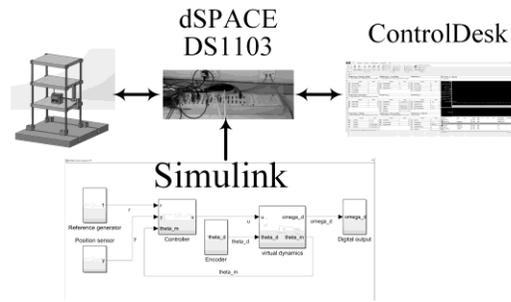

Figure 2: A photo of the mechanical testbed (a) and the flow diagram of the experiment study case (b).

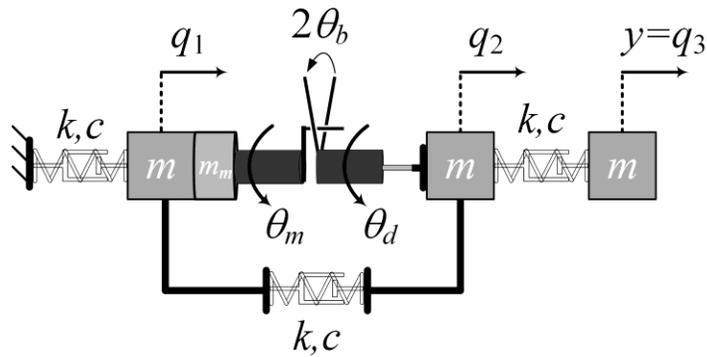

Figure 3: The mechanical system lumped-elements model.

In Section 2.1, a reduced-order lumped element model of the three-platform structure is derived, and in Section 2.2, an experimental model of the system is identified and calibrated to the analytical model.

### 2.1. Lumped element model

To obtain the lumped-elements model of the experimental system, shown in Figure 3, it is assumed that the plates' masses are equal and given by $m$, that the beams' masses are negligible, the equivalent stiffness of each of the beams, $k_b$, is linear and the same, so that the total stiffness is $k = 4k_b$. Proportional damping is assumed, and the effect of gravity is neglected. The bottom middle and top plates displacement from rest are denoted as $q_1, q_2,$ and $q_3$, respectively. The motor is initially assumed to be free of backlash and rigid, so the motor and lead screw angles, denoted as $\theta_m$ and $\theta_d$, respectively, are same. The following coordinate transformation is used from the first two plates plates' absolute displacement to their relative motion:

$$v = q_2 - q_1 \leftrightarrow q_1 = q_2 - v \tag{1}$$

Such that their relative motion $v$, is proportional to the lead screw angle:

$$v = p\theta_d \tag{2}$$

where $p$ denotes the lead screw pitch. The relative motion of the first two plates, $v$, is the assumed input to the linear system and the displacement of the top mass, $q_3$, is taken as the overall output of the control system, $y$.

Based on these assumptions and by denoting the stepper motor's mass as $m_m$, and its torque $T_m$, one can write the kinetic energy, $T$, the potential energy, $V$, the Rayleigh's dissipation function, $D$, and the stepper motor's virtual work, $\delta W$ as:

$$T = \tfrac{1}{2}(m + m_m)\dot{q}_1^2 + \tfrac{1}{2}m\dot{q}_2^2 + \tfrac{1}{2}m\dot{q}_3^2 \tag{3}$$

$$V = \tfrac{1}{2}kq_1^2 + \tfrac{1}{2}k(q_2 - q_1)^2 + \tfrac{1}{2}k(q_3 - q_2)^2 \tag{4}$$

$$D = \tfrac{1}{2}c\dot{q}_1^2 + \tfrac{1}{2}c(\dot{q}_2 - \dot{q}_1)^2 + \tfrac{1}{2}c(\dot{q}_3 - \dot{q}_2)^2 \tag{5}$$

$$\delta W = T_m \delta\theta_m = T_m \delta\theta_d = pT_m \delta v \tag{6}$$

and substitute Eq. (1) into Equations (3)-(5) yields:

$$T = \tfrac{1}{2}(m + m_m)(\dot{q}_2 - \dot{v})^2 + \tfrac{1}{2}m\dot{q}_2^2 + \tfrac{1}{2}m\dot{q}_3^2 \tag{7}$$

$$V = \tfrac{1}{2}k(q_2 - v)^2 + \tfrac{1}{2}kv^2 + \tfrac{1}{2}k(q_3 - q_2)^2 \tag{8}$$

$$D = \tfrac{1}{2}c(\dot{q}_2 - \dot{v})^2 + \tfrac{1}{2}c\dot{v}^2 + \tfrac{1}{2}c(\dot{q}_3 - \dot{q}_2)^2 \tag{9}$$

Deriving the Euler-Lagrange matrix equation results in the following matrix equation:

$$\begin{bmatrix} m + m_m & -m - m_m & 0 \\ -m - m_m & 2m + m_m & 0 \\ 0 & 0 & m \end{bmatrix}\begin{bmatrix} \ddot{v} \\ \ddot{q}_2 \\ \ddot{q}_3 \end{bmatrix} + \begin{bmatrix} 2c & -c & 0 \\ -c & 2c & -c \\ 0 & -c & c \end{bmatrix}\begin{bmatrix} \dot{v} \\ \dot{q}_2 \\ \dot{q}_3 \end{bmatrix} + \begin{bmatrix} 2k & -k & 0 \\ -k & 2k & -k \\ 0 & -k & k \end{bmatrix}\begin{bmatrix} v \\ q_2 \\ q_3 \end{bmatrix} = \begin{bmatrix} pT_m \\ 0 \\ 0 \end{bmatrix} \tag{10}$$

The first equation relates the force needed from the stepper motor to enable the desired trajectory, while the latter two equations represent the internal dynamics of an unforced 2-degrees-of-freedom structure driven with a base excitation.

When neglecting the first equation of motion and splitting the remaining submatrices, such that all the matrices multiplying $v$ and its derivatives are on the right-hand-side, the following equations of motion are obtained:

$$\begin{bmatrix} 2m + m_m & 0 \\ 0 & m \end{bmatrix}\begin{bmatrix} \ddot{q}_2 \\ \ddot{q}_3 \end{bmatrix} + \begin{bmatrix} 2c & -c \\ -c & c \end{bmatrix}\begin{bmatrix} \dot{q}_2 \\ \dot{q}_3 \end{bmatrix} + \begin{bmatrix} 2k & -k \\ -k & k \end{bmatrix}\begin{bmatrix} q_2 \\ q_3 \end{bmatrix} = \begin{bmatrix} m + m_m \\ 0 \end{bmatrix}\ddot{v} + \begin{bmatrix} c \\ 0 \end{bmatrix}\dot{v} + \begin{bmatrix} k \\ 0 \end{bmatrix}v \tag{11}$$

Applying Laplace's transformation, under zero-initial condition assumption, gives the transfer matrix that connects the generalized base excitation, $v$, to the individual platforms' displacements:

$$\begin{bmatrix} q_2 \\ q_3 \end{bmatrix}(s) = \begin{bmatrix} G_{v \to q_2} \\ G_{v \to q_3} \end{bmatrix} v(s) = \frac{(m+m_m)s^2 + cs + k}{\Delta_G}\begin{bmatrix} (ms^2 + cs + k) \\ (cs + k) \end{bmatrix} v(s) \tag{12}$$

$$\Delta_G = (2m + m_m)ms^4 + (4mc + m_m c)s^3 + (4mk + m_m k + c^2)s^2 + 2cks + k^2 \tag{13}$$

In the rest of this paper, we are mainly interested in the effect of $v$ on the top platform position; therefore, the following notation is used:

$$G_2 = G_{v \to q_3} \tag{14}$$

## 2.2. Experimentally obtained model

A stepped sine identification technique was used to obtain the experimental frequency response function (FRF) of $G_2$. At each test, a single-frequency sinusoidal signal was used to excite the stepper motor for 30 seconds. The last 10 seconds of the responses truncated Fourier's series and were fitted using a linear least-squares and the following model:

$$y(t) = A_{ab} \begin{bmatrix} a_c \\ b_s \\ a_0 \end{bmatrix} \tag{15}$$

where

$$A_{ab} = [\cos(\omega_0 t) \quad \cdots \quad \cos(N\omega_0 t) \quad \sin(\omega_0 t) \quad \cdots \quad \sin(N\omega_0 t) \quad 1] \tag{16}$$

and $\omega_0 = 2\pi f_0$, $f_0$ being the excitation frequency. $a_c^T = [a_1 \quad \cdots \quad a_N]$ denotes the cosine coefficient vector, $b_s^T = [b_1 \quad \cdots \quad b_N]$ denotes the sine coefficient vector, and $a_0$ denotes the constant, dc, coefficient. The coefficient vector can be estimated in the least-squares-sense by populating Eq. (15) with the measurement, where each consecutive measure is written in a new row by:

$$\begin{bmatrix} a_c \\ b_s \\ a_0 \end{bmatrix} = A_{ab}^{\dagger} y \tag{17}$$

where $A_{ab}^{\dagger}$ denotes the generalized left inverse of $A_{ab}$, and is given by:

$$A_{ab}^{\dagger} = \left(A_{ab}^T A_{ab}\right)^{-1} A_{ab}^T \tag{18}$$

The experimental FRF is built up one frequency at a time by measuring the motor's angle and the top platform's position for the sinusoidal wave excitation at that frequency. Then the steady-state responses are used to populate Eq. (15), and each of the signals' complex-valued Fourier's coefficient at the excitation frequency is derived from the fitted values of Eq. (17) as:

$$c^y(f_0) = \frac{1}{2}\left(a_1^y - ib_1^y\right) \tag{19}$$

where $y$ is either $\theta$ or $q_3$, and $i = \sqrt{-1}$ denotes the imaginary unit.

Then the FRF is computed at the excitation frequency by dividing the output by the input complex Fourier's coefficient:

$$G_2(f_0) = \frac{c^{q_3}(f_0)}{c^u(f_0)p} = \frac{a_1^{q_3} - ib_1^{q_3}}{(a_1^u - ib_1^u)p} \tag{20}$$

Where the input is first multiplied by the lead-screw pitch, $p$, to transform $\theta_m$ to $v$, according to Eq. (2).

After obtaining the experimental FRF at frequencies between 0.5-15 Hz, the parameters of the analytical model of Eq. (12) were fitted to the experimentally obtained $G_2$ using the nonlinear least squares method with a cost function that aim to minimize the logarithmic absolute value of the difference between the measured FRF and the analytical one obtained by substituting the values in Eq. (12). Also, a general rational transfer function model with four poles and three zeros was fitted using a least squares approach (the invfreqs function in MATLAB with unity weights). The identified system lumped parameters are: $m=16.06$ kg, $k = 43.57$ N/mm, $m_m=2.4$ kg, and $c = 0.012$ N/mm/s.

Figure 4 shows the top platform's FRF obtained experimentally, the FRF of the fitted rational transfer function model, and that of the lumped element analytical model. For the analytical model, there is a significant shift in the second mode's frequency and amplitude and a slight shift in the frequency of the zero. For the fitted rational transfer function, the second resonance is captured, but there is a significant shift in the frequency of the zero. Moreover, upon inspecting the zeros of the fitted rational transfer function, it is found that there is a right-hand-side zero, meaning that the modelled system is not minimum-phase, as is expected from the physics of the real system, so the analytic lumped element model was chosen to model $G_2$ below. However, both models capture the first mode's natural frequency and modal damping. The facts that all models are not exact and that a model-based controller must be robust are at the heart of modern control and will be emphasised when considering the controller structure, which is based on the analytical model since it has a stronger connection to the underline physics.

## 3. Backlash

### 3.1. Backlash model

The friction-driven hysteresis model [4,33] is adopted for the backlash. This simplified model is chosen to reduce the complexity of the more detailed backlash models while keeping the nonsmooth nonlinear behaviour of the backlash, which is most likely to affect the feedback significantly. The backlash mechanism in the lead screw connecting the motor to the structure can be modelled as a dead zone for the angular velocity. Each time the rotation direction changes, i.e., the motor and lead screw angular velocities cross the zero, a new dead zone of a given backlash gap angle begins.

The backlash nonlinear operator is given by

$$\omega_d = g(z, \sigma)\omega_m \tag{21}$$

where $z = \theta_m - \theta_d$ denotes the difference between the motor and lead screw angles, $\theta_m$ and $\theta_d$, $\omega_m$ and $\omega_d$ are the angular velocities of the motor and lead screw, respectively, $\sigma = \text{sign}(\omega_m)$ denotes the motor's direction of rotation. Denoting the backlash gap angle, or half-width, as $\theta_b$, $g(z, \sigma)$ is given by:

$$g(z, \sigma) = \begin{cases} 1 & (\sigma > 0 \ \& \ z \geq \theta_b) | (\sigma < 0 \ \& \ z \leq -\theta_b) \\ 0 & \text{Otherwise} \end{cases} \tag{22}$$

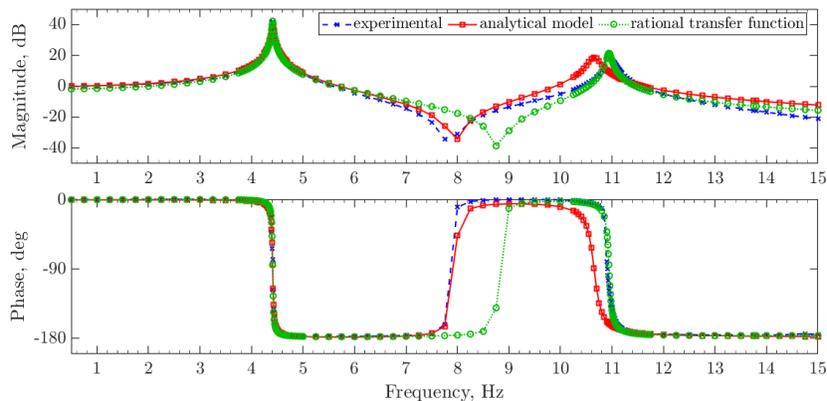

Figure 4: $G_2$ frequency response function.

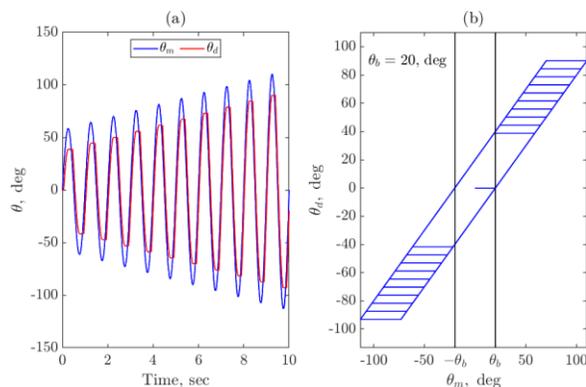

Figure 5: Input and output of the backlash model for a sinusoidal input with a growing amplitude versus time (a) and in state-space (b).

Figure 5 shows the output of the backlash for $\theta_b = 20$ deg, and a sinusoidal input with a growing amplitude: $\theta_m(t) = (1 + t/10)\sin(2\pi t)$. In Figure 5 the effect of the velocity dead zone on the lead screw's angle is presented. Each time the direction of rotation changes, $\theta_d$ remains constant until the backlash is transverse. Then, $\theta_d$ follows $\theta_m$ with a constant lag of $\theta_b$. The former is best seen in Figure 5(a), while the latter is in Figure 5(b).

### 3.2. Virtual backlash control

It is desirable to introduce virtual backlash into the system in a controllable manner to allow non-destructive evaluation of different control schemes without waiting for progressive wear and to investigate their robustness to changes in the backlash gap angle.

The virtual backlash is introduced here by employing the stepper motor driver, which controls the stepper motor pulse rate (angular velocity) and introducing a virtual dc-motor that is realized using the dSPACE system. The virtual backlash is situated between the virtual dc-motor and the stepper motor and is realised using a nonlinear control law based on Eq. (22).

The nonlinear virtual backlash block diagram is shown in Figure 6, where $G_1$ represents the stepper motor, $G_2$ the three-platform structure, $C$ the controller to be designed, $G_v$ the virtual motor, and $g$ the backlash nonlinear operator of Eq. (22). Note that $\omega_m$, $\sigma$, and $\theta_m$ are all internal, virtual, signals and $\theta_d$ can be obtained to a relative good accuracy using the stepper motor's encoder. The virtual motor dynamics was chosen to be linear, with the following transfer-function model:

$$G_v = \frac{k_t}{J_v L s^2 + J_v R s + k_b k_t} \quad (23)$$

where $J_v = 5$ kg·mm² is the virtual motor's inertia, $L = 10$ mH and $R = 44.72$ ohm are the virtual inductor and resistor values, respectively, and $k_t = k_b = 0.5$ denote the virtual motor's torque and back-emf constants. The values were chosen to represent a real dc-motor. However, $G_v$ does not have to be linear or time-invariant. For example, time-varying input and output saturations could have also been added.

#### 3.2.1. Verification of backlash hybrid testing

To verify the proposed nonlinear control law of the virtual backlash, the system was tested experimentally for different virtual backlash angle values. Since the experimental system is unstable if no position controller is used, a PID controller was initially used, tuned for the virtual and physical interconnected systems with no backlash. The PID controller was designed using the MATLAB PID Tuner app [34]. Based on the tuning procedure, a parallel filtered PID architecture was chosen with the controller's parameter set to: $K_p = 1.59, K_i = 1.01, K_d = -0.56$, and $T_f = 0.35$.

A numerical simulation was also used to verify the experimentally obtained results. In the numerical simulation, the discrete nature of the stepper motor and any additional dynamics were neglected, and the motor was modelled as a pure integrator with a gain adjusted to reflect the actual pulse per revolution value.

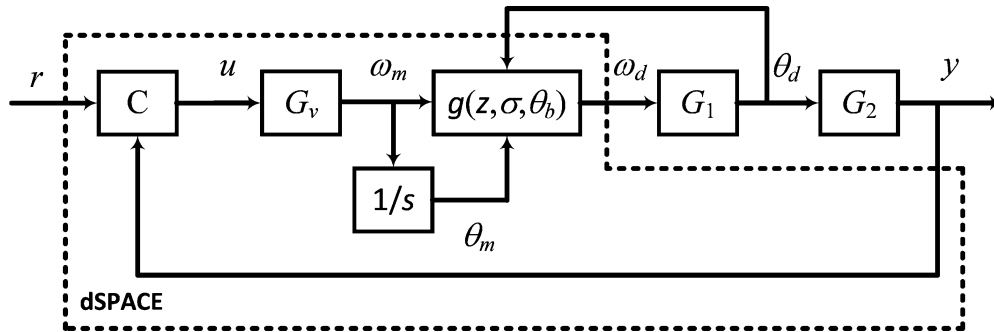

Figure 6: Virtual backlash block diagram

Figure 7 shows the experimentally and numerically obtained step responses for different amounts of backlash introduced into the system virtually when the system is being controlled using a PID controller. Note that when there is no backlash, Figure 7(a), there is an excellent agreement between the experiment and simulation. Also, for the no backlash case, the PID stabilises the otherwise unstable system but with some overshoot and a long settling time. This is because the structural system damping is rather small, which imposes difficulties when using a low-order controller like the PID.

When the backlash is introduced, Figure 7(b-d), the experiment and simulation both exhibit a loss of stability in the form of limit cycle oscillations. But slightly different periods of oscillations are observed, causing a slight phase difference between the measured and simulated results. The differences between the results are due to both mismodelling of the linear plant and the existence of unmodelled dynamics. The amplitude of the limit cycles increases as the backlash gap angle increases.

Figure 8 shows another representation of the step responses data, plotting the stepper motor's angle versus that of the virtual motor. Note that in all cases where the backlash is present, the expected curve is seen, similar to the one obtained in the simulation of Section 3.1. In addition, there is an excellent agreement between the predicted width of the simulation and the ones measured experimentally. By measuring the width, it is possible to conclude that the proposed approach was indeed able to introduce the desired amount of backlash into the experimental system. It thus provides a good platform to compare the model-free PID controller with a model-based control strategy.

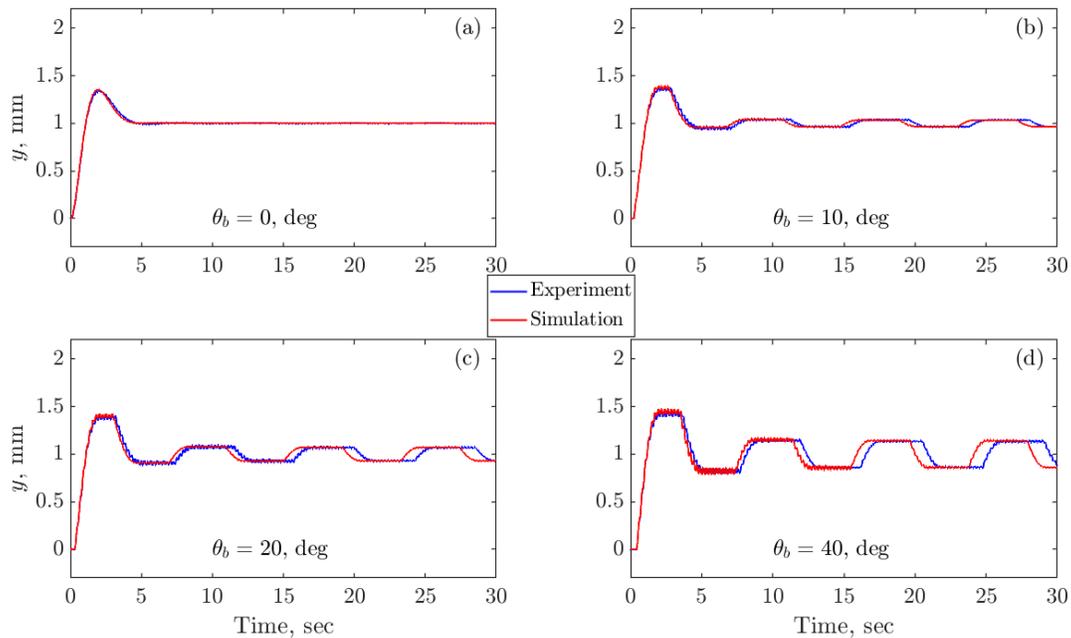

Figure 7: Virtual backlash verification experiments. Experiment and simulation step response, showing the output versus time.

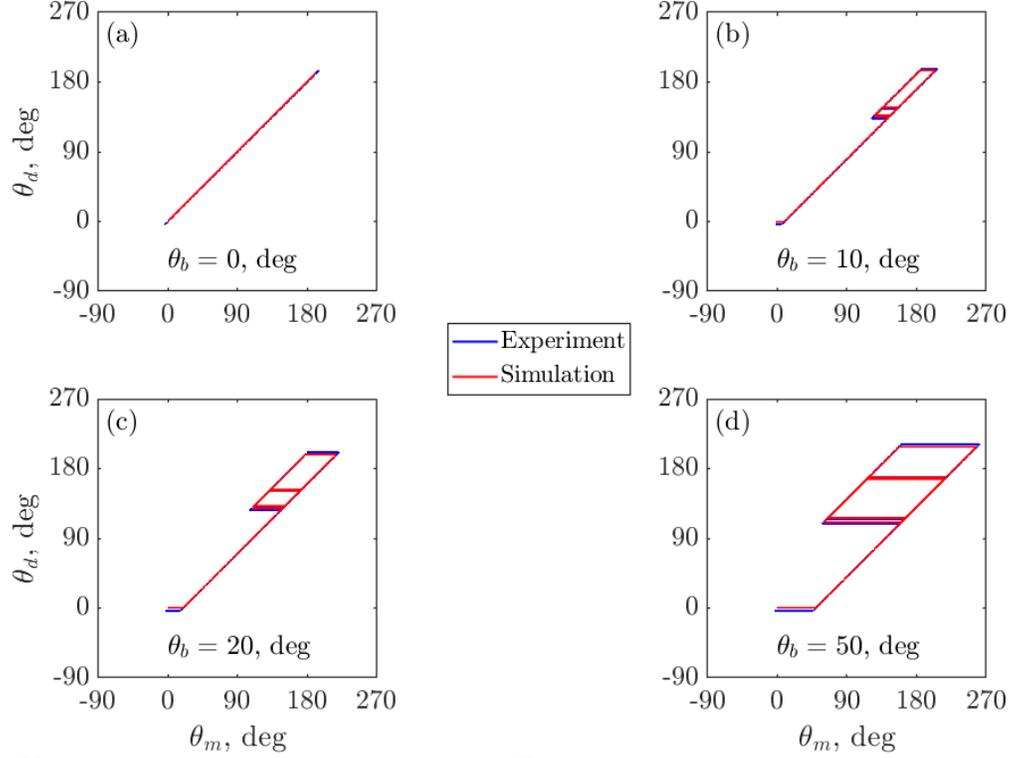

Figure 8: Virtual backlash verification experiments. Experiment and simulation step response, showing the stepper motor's angle versus the virtual motor's angle.

## 4. Linear Internal model control design

The proposed modified linear IMC architecture is shown in Figure 9, where $G_1$ and $G_2$ represents the plant responses of the motor and structure, and $\hat{G}_1$ and $\hat{G}_2$ are the internal models of these plant responses. The controller inputs are the reference command signal, $r$, motor's angle, $\theta_m$, and top platform position, $y$. Since $G_1$ is not stable an inner gain-feedback loop is closed around it, where the constant $K_\theta$ denotes the inner loop gain. The internal model is used in the middle branch to obtain the estimated output $\hat{y}$, while the feedback is closed on the model estimation error $\hat{\varepsilon}$. $\theta_r$ is a virtual inner signal of the controller used as the reference input of the inner loop. $W$ is the prefilter to be designed such that $\hat{y}$ satisfies the desired tracking requirements.

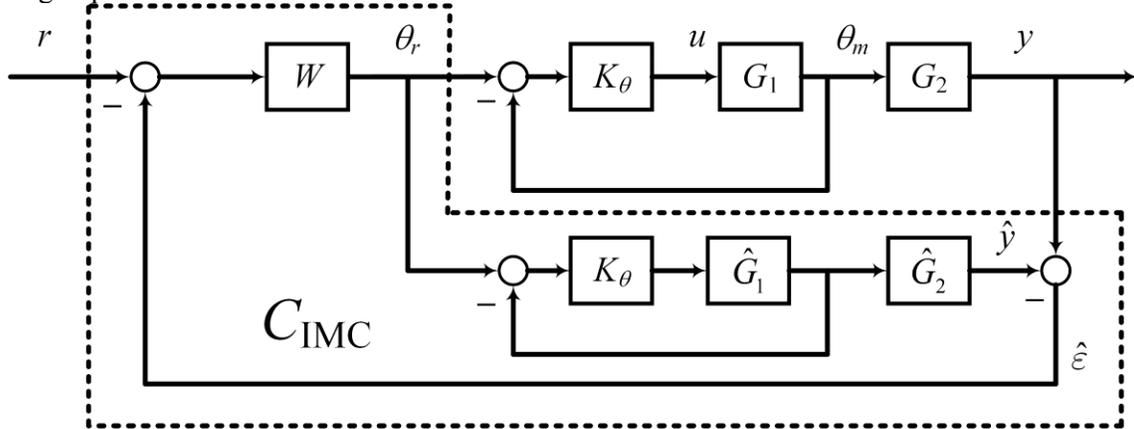

Figure 9: Linear IMC block diagram, where the IMC controller is contained within the dashed line.

### 4.1. Prefilter design

The classical design approach for the prefilter employs a feedforward reference model. Assuming that the nominal plant model, $\hat{G}$ in Figure 1, is linear, stable and minimum phase and denoting its poles excess as $n - m$, one can choose as the reference model, $G_r$, any stable transfer function with a larger or equal pole excess. The minimum phase assumption can be relaxed by including all nonminimum phase zeros of the nominal model in the reference model or alternatively by first decomposing the nominal plant using the all-pass minimum phase decomposition [14], but this was not required here since the underline physics is minimum-phase.

For the design of the proposed IMC of Figure 9, $\hat{G}_2$ was assumed to be given by the lumped element model estimate of Section 2.2. It is assumed that $\hat{G}_1$, which represent the virtual motor, is known accurately from the virtual motor controller in Section 3.2. As suggested in [22], the motor's dynamic was not inverted, and the prefilter was designed only based on $\hat{G}_2$. The prefilter structure is thus given by:

$$W = G_r \hat{G}_2^{-1} \qquad (24)$$

Denoting the inner feedback and its model by $G_\theta$ and $\hat{G}_\theta$ respectively, where:

$$G_\theta = K_\theta G_1 / (1 + K_\theta G_1) \qquad (25)$$

and $\hat{G}_\theta$ is derived the same way by replacing $G_1$ with its model $\hat{G}_1$, the equivalent outer feedback controller, from the tracking error, $e = r - y$, to the reference angle $\theta_r$, in this case, can be written as:

$$C_e = \frac{G_r \hat{G}_2^{-1}}{1 - G_r \hat{G}_\theta} \qquad (26)$$

### 4.2. Linear robust stability and performance

The case where $G_2$ is related to its internal model $\hat{G}_2$ by a multiplicative unstructured uncertainty is considered, such that:

$$G_2(\omega) = \hat{G}_2(\omega)(1 + \Delta(\omega)) \qquad (27)$$

where $|\Delta(\omega)|$ is the multiplicative uncertainty radius, which is frequency dependent.

For the nominal case, when there is no uncertainty, $\Delta(\omega) \equiv 0$, the designed IMC feedback loop is stable for any choice of a stable $G_r$. Therefore, based on the small gain theorem [14,15], robust stability is ensured if the nominal complementary sensitivity function, $T_0$, satisfies:

$$|T_0(\omega)| < 1/|\Delta(\omega)| \qquad (28)$$

It follows that the nominal complementary sensitivity function for the IMC controller of Figure 9 is equal to:

$$|T_0(\omega)| = |G_r(\omega) G_\theta(\omega)| \qquad (29)$$

Based on the relation between the complementary sensitivity and the chosen reference model, one can tune the reference model, $G_r$, and the inner feedback gain, $K_\theta$, such that $T_0$ is kept small at frequencies when the uncertainty radius, $|\Delta(\omega)|$, is large to ensure the robust stability of the closed loop.

Furthermore, when considering the robust performance, it is required that the sensitivity function $S(\omega)$ will be small at the desired bandwidth. For the IMC controller of Figure 9, the sensitivity function is given by:

$$|S(\omega)| = \left| \frac{1 - G_r(\omega) G_\theta(\omega)}{1 + G_r(\omega) \hat{G}_\theta(\omega) \Delta(\omega)} \right| \qquad (30)$$

So robust performance is maintained at frequencies where $T_0(\omega) = G_r(\omega) G_\theta(\omega) \approx 1$, as long as the robust stability criterion is maintained. These two requirements can be readily obtained when considering the step response, i.e., $\omega \to 0$.

### 4.3. Experimental results and comparison to a model-free controller (PID)

Based on the discussion of Section 4.2, it is noted that the reduced-order model obtained experimentally predicts the system's behaviour accurately near the first resonance but less accurately near the second resonance frequency, as shown in Figure 4. To ensure that the nominal complementary sensitivity is small near the second resonance frequency, the motor gain in Figure 9, $K_\theta$, was tuned using the root locus of $\hat{G}_1$ and was chosen to be equal to 10, and the reference model, $G_r$, was chosen as a second-order low-pass of the form:

$$G_r = \left(\frac{5.834}{s+5.834}\right)^2 \tag{31}$$

The prefilter $W$ is then derived using Eq. (24) and discretised using the zero-order assumption for a sampling time of 1 millisecond. It is worth noting that the resulting sampling frequency is two orders of magnitude larger than the natural frequency of the second structural resonance.

Figure 10 shows the resulting experimental step response of the IMC controller, designed in Section 4.1, compared with the measured response with the PID controller, designed in Section 3.2.1, in the case with no added backlash. It is noted that although the peak control voltage for the IMC controller is slightly higher than for the PID, the integral control effort is similar, and the IMC controller outperforms the PID, converging to the steady-state in under a second with no overshoot. The results of Figure 10 also show that the designed IMC controller is robust to the modelling uncertainties.

It is noticeable that the control signal for the IMC controller contains oscillations, which are at the frequency of the zero in the frequency response of the plant model. These oscillations are present since the frequency of this zero is not exactly the same as that of the zero in the physical plant response, and so the controller enhances the measured vibrations at this frequency since it inverts the frequency response of the assumed plant response in the internal model, $\hat{G}_2$.

Next, the step responses of the same linear IMC and PID controllers were compared for different virtual backlash gap angles.

The results are shown in Figure 11, which shows that while the PID has severe performance degradation and can lose its stability for an increasing amount of backlash, the IMC maintain stability and good transient performance even for a significant amount of backlash. Some residual vibrations do begin to occur as the backlash gap angle increases.

Upon further inspection of the residual vibrations, it is noted that they occur at similar frequencies, regardless of the amount of backlash. Therefore, a single-mode damped vibration model of the form

$$\eta = y_{res} - r = A_{res}e^{-2\pi f_r \zeta_r (t-t_s)} \sin\left(2\pi f_r \sqrt{1-\zeta_r}t + \varphi\right) \tag{32}$$

was fitted to the steady-state measurements. The identified values are $f_r = 4.44Hz$ and $\zeta_r = 0.003$, which are similar to the ones of the model's first mode, which can be computed from the roots of Eq. (13) after substituting the identified structure parameters, as $f_1 = 4.40\text{Hz}$ and $\zeta_1 = 0.0038$. Therefore, it can be concluded that the residual vibrations period is the same as the first structural mode damped frequency, but their amplitude, $A_{res}$, depends on the amount of backlash. This supports the finding of [22] and could be used to design a frequency lock-in filter to assess the amount of backlash in the system.

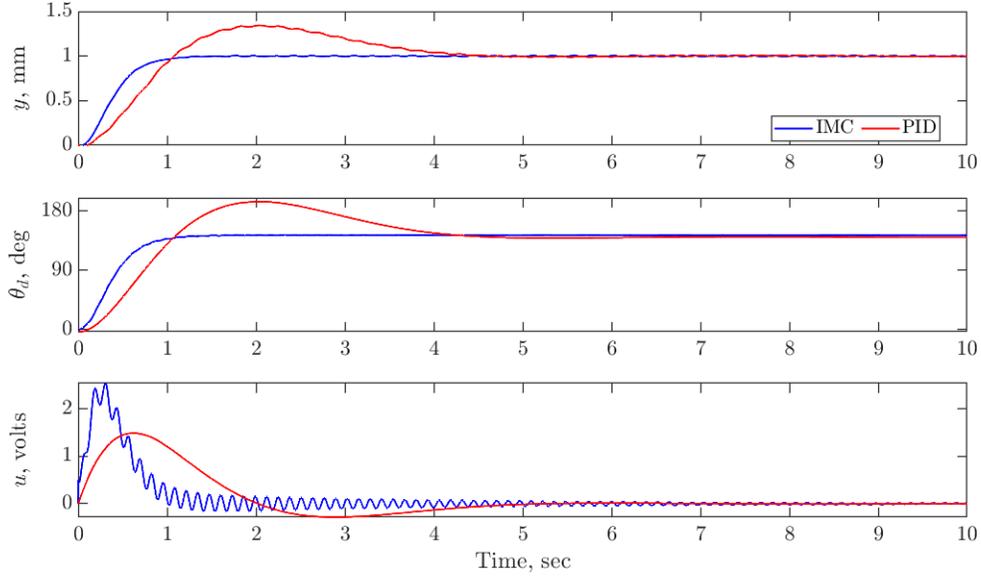

Figure 10: Measured closed-loop step response of the IMC and PID controllers with no added backlash

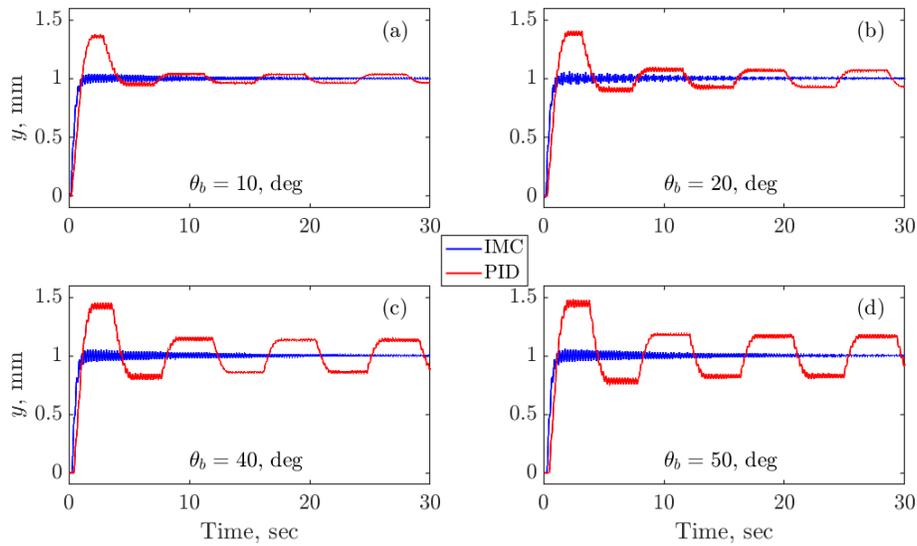

Figure 11: Measured closed-loop step response of the linear IMC and PID controllers with virtual backlash

### 4.4. The effect of different command signals and growing backlash

The virtual backlash gap angle was gradually increased during testing to investigate the linear IMC's robustness further. The backlash gap angle was increased by 10 degrees each 30 seconds for values between 0 and 200 degrees. The reference input signal used in this experiment was a square-wave signal with a 1-millimetre amplitude and a period of 60 seconds. The backlash increase was synchronised to be approximately 15 seconds after each change of the reference signal.

Figure 12 shows the experiment results. Stability is again always maintained, even for large backlash gap angles. However, the transient behaviour starts to deteriorate when the backlash becomes significant, and an overshoot is then observed. The response converges to the desired steady-state values, but the virtual motor angle drifts inside the backlash at steady-state. Therefore, adding a small dead zone into the controller might prove advantageous.

In reality, the command signal to be followed does not have to be a square wave or piecewise constant and may be obtained as a smooth spline or sinusoidal wave. Therefore, the response of the designed IMC-based controller when a smooth command signal is used was checked experimentally, as the backlash was again gradually increased, as above. The input frequency was chosen to be slow enough for accurate tracking when no backlash is introduced.

Figure 13 shows the experimentally obtained results for a sinusoidal command signal with a frequency of 0.03 Hz when the backlash gap angle is increased by 10 degrees every 30 seconds. It is noted that initially, the tracking is excellent, with a small delay and gain, which are not noticeable in the figure timescale. The control signal is also smooth and sinusoidal, as expected. Then when the backlash gap angle increases, the output signal develops a sharp corner near each peak or direction change. These corners are the result of the phase delay between the angles of the motor and the lead screw. The motor angle first needs to transverse the gap, and only then the lead screw start to follow.

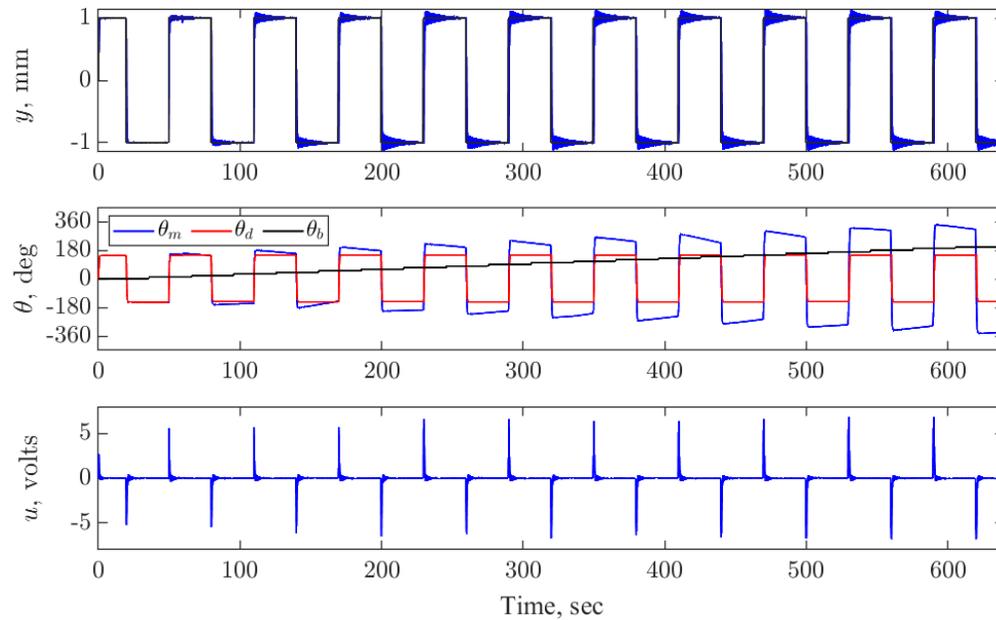

Figure 12: Measured closed-loop response of the linear IMC controller for a growing backlash and a square-wave input

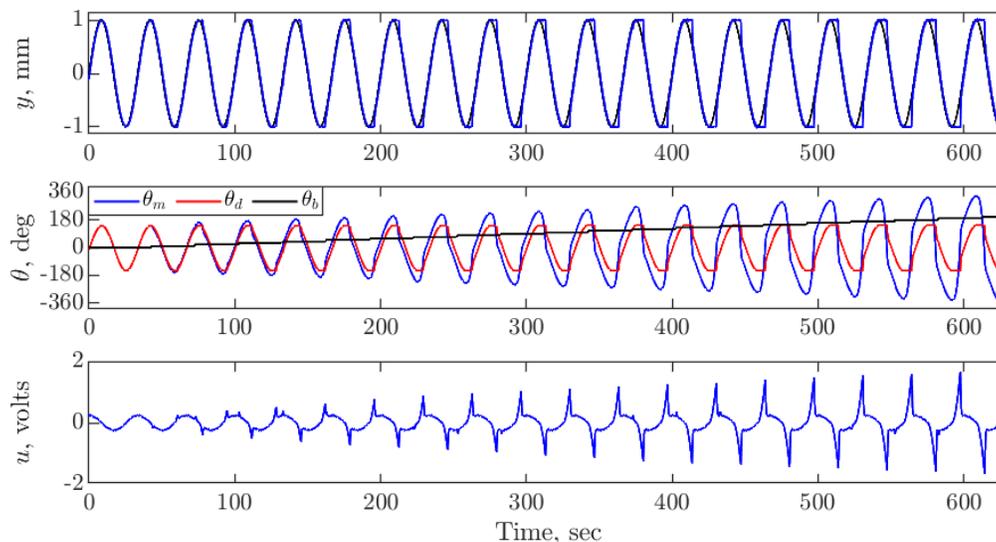

Figure 13: Measured closed-loop response of the linear IMC controller for a growing backlash and a sinusoidal command signal with 0.03 Hertz frequency

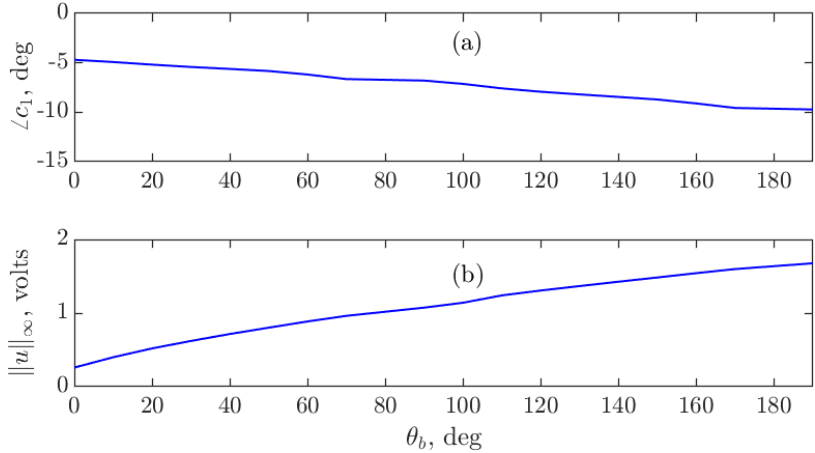

Figure 14: The effect of increasing the backlash gap angle on the phase shift between the output's principal harmonic and command signal (a) and on the control signal peak (b) based on the measurements of Figure 13.

The fundamental component of the output was estimated using the least-squares method, and its phase shift compared to the command signal is plotted in Figure 14(a) as a function of the backlash gap angle at the beginning of each new cycle of the reference command signal. It can be seen that the phase shift, which is proportional to a time delay, grows linearly with the backlash angle.

The control signal also becomes more pronounced at the direction changes when the backlash increases; this corresponds to the need to make the motor transverse the backlash faster and then to account for the error between the command and output signals. When the tracking error is diminished, the control signal returns to follow the sinusoidal profile of the linear case, resulting in an abrupt decrease in its absolute value. This behaviour results in noticeable peaks in the control signals. Figure 14(b) shows the amplitude of the control signal's peaks versus the virtual backlash gap angle. The peak magnitude seems to be proportional to the backlash gap angle. Thus, if the amount of backlash in the system needs to be identified, it might also prove helpful to observe the control signal profile.

### 4.5. Nonlinear stability analysis using the describing function analysis

As seen in Section 4.3, the backlash can cause stability loss in feedback systems, which can be approximately analysed using the describing function method [35]. The describing function of the backlash was formulated based on the model of Section 3.1 and is derived in the appendix as:

$$N(\chi) = \frac{1}{\pi}\left(\frac{\pi}{2} + \gamma + 2(1-2\chi)\sqrt{\chi(1-\chi)}\right) + i\frac{4}{\pi}\chi(\chi-1) \tag{33}$$

where $\chi \equiv \theta_b/\Theta_m \in (0,1)$, and $\Theta_m$ denotes the amplitude of the motor's angle, which is the input to the backlash operator. Note that $|N(\chi = 0)| = 1$, and as $\chi$ grows, it is monotonic decreasing, with an asymptotic value of 0 at $\chi = 0$. Also $\angle N(0) = 0$ degrees, and as $\chi$ grows, it is monotonic decreasing, with an asymptotic value of $-90$ degrees. The dependency of the describing function on the frequency is not direct. Instead, they are connected via the underline linear system frequency response, which will dictate the input amplitude of the backlash's describing function $\Theta_m$.

To analyse the stability, the modified Nyquist criterion is used. The criterion states that stability loss will occur if the following equation is satisfied:

$$G_{\text{OL}}(s = i2\pi f) = -\frac{1}{N(\chi)} \tag{34}$$

where $G_{\text{OL}}$ denotes the open-loop transfer function. For the two designed controllers $G_{\text{OL}}$ is given by:

$$G_{\text{OL}}(C_{\text{IMC}}) = \frac{W}{1-W\hat{G}_\theta G_2} G_\theta G_2, \text{ and } G_{\text{OL}}(C_{\text{PID}}) = \left(K_p + \frac{1}{s}K_i + \frac{s}{\tau_f s+1}K_d\right)G_1 G_2 \tag{35}$$

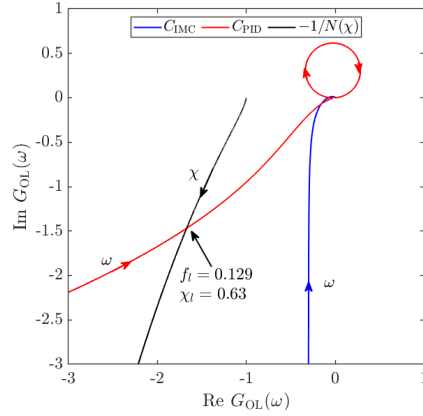

Figure 15: Nonlinear stability analysis for the linear PID and IMC controllers

If the two complex loci of Eq. (34) intersect, the stability will be lost, and instead, a limit cycle will appear.

Figure 15 shows the results of this analysis for the two designed controllers and the backlash describing function. It is noted that when using the PID controller, even a small amount of backlash will result in stability loss and the appearance of the limit cycle at the frequency of $f_l$ and at an amplitude ratio of $\chi_l$. Consequently, the first harmonic's amplitude of the output signal will be:

$$A_l = |G_2(f_l)N(\chi_l)|\Theta_m \qquad (36)$$

On the contrary, if using the IMC controller, the two curves are well separated. Therefore, the stability will be kept for any amount of backlash when using the IMC-based controller.

The measured frequency of the limit cycle generated by the PID controller in Figure 11 is within 9% of the prediction by this analysis. The amplitude of oscillation is less accurately predicted due to model mismatch but is within 20% of the predicted one.

## 5. Nonlinear IMC controller

As seen in Section 4.3, when the backlash gap angle becomes larger, the amplitude of the residual vibration and the control signal also increases. One of the effects of the backlash, seen in Section 4.4, is that the controller attempts to minimise the vibration by changing the motor's angle when the mechanism is inside the gap. However, the actual lead screw's angle is not affected, resulting in a drift of the motor's angle. A dead zone was added in the inner motor's feedback loop to avoid this so-called drift. Also, as in [22], a backlash estimator was added to the inner feedback that aims to control the estimated lead screw's angle instead of the motor. If either the dead zone, the backlash estimator, or both are added to the inner feedback, they must also be added to the internal model branch. The proposed structure of the nonlinear IMC controller is then shown in Figure 16. The modelled system used to obtain the predicted output, $\hat{y}$, is also updated to represent the backlash between the motor and lead screw. The effects of the nonlinear elements in the controller were investigated experimentally. Section 5.1 report the results of using only the dead zone, and Section 5.2 those when using only the backlash estimator.

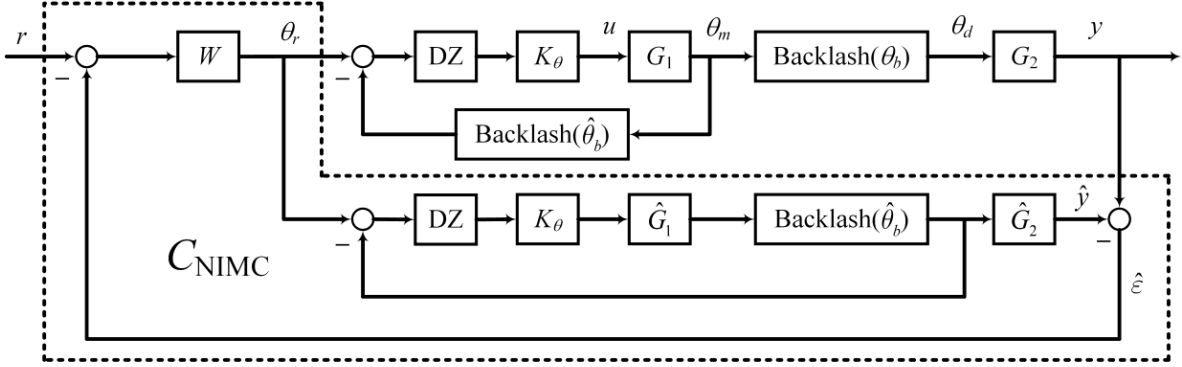

Figure 16: Block diagram of the nonlinear IMC architecture. A dead zone (DZ) and a sampled-data model of the backlash are added to the internal feedback loop to compensate for the lead screw backlash. $\theta_b$ and $\hat{\theta}_b$ denote the actual and estimated backlash gap angle, respectively.

**5.1. Dead Zone Effect**

To avoid the observed drift of the motor's angle inside the backlash gap and minimise the effect of measurement noise, a dead zone was added to the angle inner-feedback loop. However, the presence of the dead zone allows a small bias from the desired reference. Therefore, it is desirable to keep the dead zone small.

Figure 17 and Figure 18 show the experimental results when considering a dead zone with an angle, denoted as $\theta_{dz}$, of 0, 0.5, and 1 degrees, and different amount of backlash gap angles. Figure 17 shows the measured output signal, and Figure 18 the output control signal.

It is noted that when there is no backlash, Figure 17(a) and Figure 18(a), the addition of the dead zone also results in a less oscillatory control signal near the steady-state (Figure 18(a)) and can reduce the effect of the measurement noise at steady-state (Figure 17(a)). When a backlash is added, subplots (b),(c), and (d), the dead zone also helps to decrease the residual vibration's amplitude when tuned correctly.

Based on the experimental results, using a small dead zone in the inner feedback loop seems advantageous. A range of $\theta_{dz}$ between 0.5 to 1 degree gives a reasonable trade-off between the desire to minimise the effect of measurement noise and the steady-state bias.

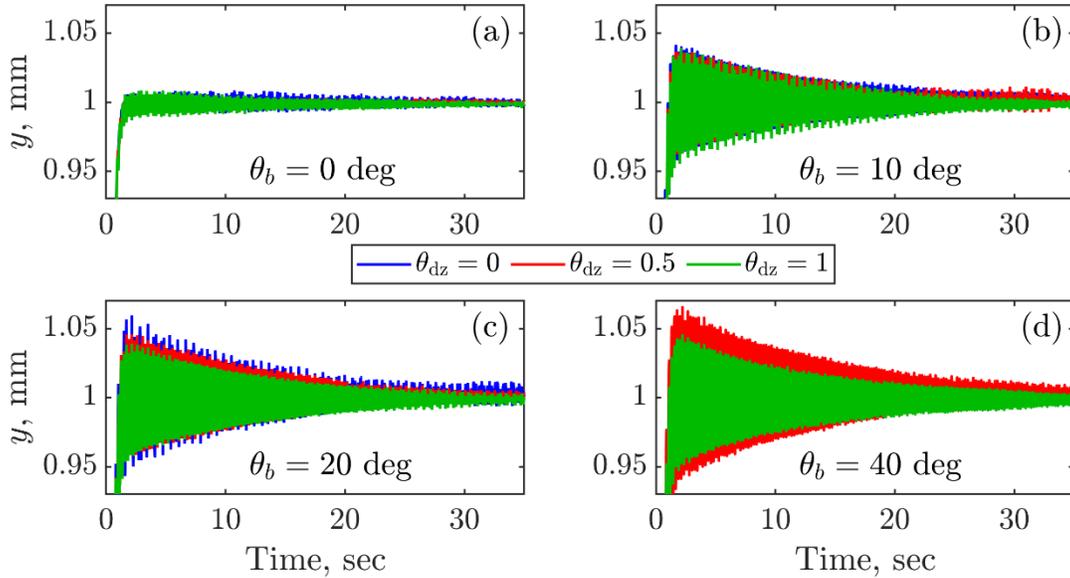

Figure 17: Step response output signal of the IMC with different dead zone angles and virtual backlash.

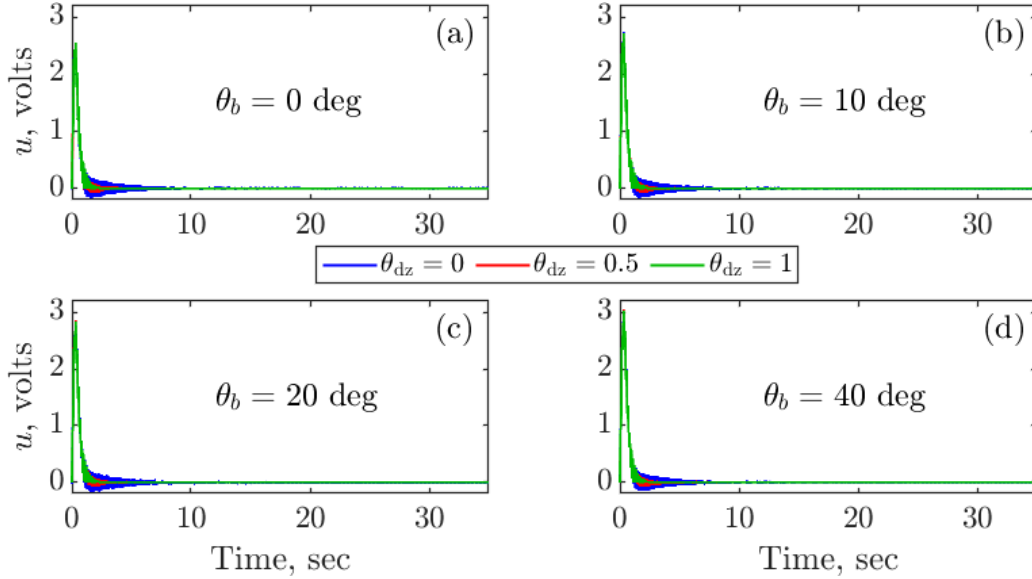

Figure 18: Step response control signal of the IMC with different dead zone angles and virtual backlash

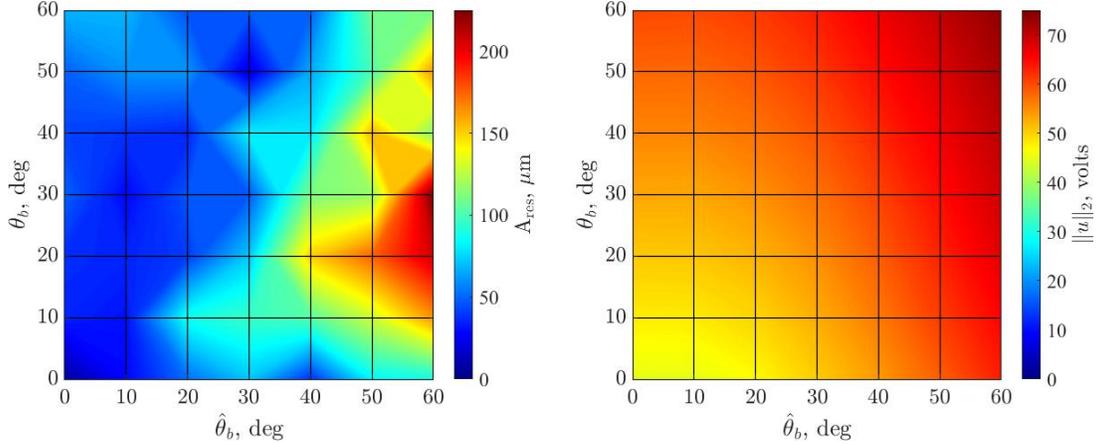

Figure 19: Backlash estimator nonlinear IMC step response for different actual and estimated backlash angles. The fitted residual vibration's amplitude is shown on the left, and the control signal norm is on the right.

### 5.2. Nonlinear Backlash Estimator

The effect of adding the backlash estimator in the inner feedback loop and in the internal model in Figure 16 was examined experimentally. Step response experiments were conducted for a grid of actual and estimated backlash angles. For each experiment, the steady-state signal was fitted using Eq. (32) to estimate $A_{\text{res}}$. In addition, the maximal voltage of the control signal was recorded.

Figure 19 shows a graphical representation of $A_{\text{res}}$ on the left and the control signal norm on the right. It can be seen that in some cases, it can be advantageous to introduce the backlash estimator to reduce $A_{\text{res}}$. However, overestimating the extent of the actual backlash results, most of the time, in an increase of $A_{\text{res}}$. Moreover, adding the backlash estimator will always increase the control signal norm compared to that of the linear IMC ($\hat{\theta}_b = 0$).

Based on the experimental results of Figure 19, it is concluded that the introduction of the backlash estimator is not necessary in this case, despite the finding of [22], which were only based on a simulation case study. The difference between the analysis done here and that of [22] is that the models used for the nominal structure in the internal model are not exact. Also, the underlined nominal structure damping in this work is an order of magnitude lower than the one used in the simulated case study of [22].

In the following sections, the sensitivity and robustness of the IMC with a small dead zone will be checked for modelling errors of the nominal plant, Section 6, and for the designed settling time, Section 7.

## 6. Robustness to changes in the nominal plant

An important feature of the IMC-based controller is its robustness to structural modifications and its sensitivity to modelling errors. Therefore, the controller's robustness to structural modifications was examined experimentally, first by adding additional physical masses on top of the middle and top platforms and then by perturbing the assumed mass and the stiffness of the internal model used in the controller design.

### 6.1. Added mass effect

To verify the robustness of the designed controller, structural modifications were introduced to the experimental system. The nominal masses were perturbed by adding 5-kilogram blocks to the different platforms. Three configurations were examined: the addition of 5 and 10 kilograms onto the top platform and 5 kilograms to both the middle and top platforms. The controller used in this set of experiments was designed using the identified nominal model of Section 2.2 with the reference model of Eq. (31) and a dead zone of $\pm 0.9$ degrees.

Figure 20(a) shows the measured step response of the output signal for the nominal case and the three configurations when there is no added backlash. No significant differences are noted between the different configurations. In fact, the second and third configurations seem to have a slightly better response than the nominal case. Again, this fact highlights that, in reality, the nominal configuration is not exact. This result again demonstrates the robustness of the IMC controller to structural modification and supports the use of a non-exact model.

An additional set of experiments were carried out with a backlash angle of 50 degrees, whose responses are shown in Figure 20(b). Similar to the case with no backlash, there are no evident changes between the configurations. To quantities the effect of the structural modification, the steady state responses were again fitted using the single mode decay model of Eq. (32), and the results are summarized in Table 1.

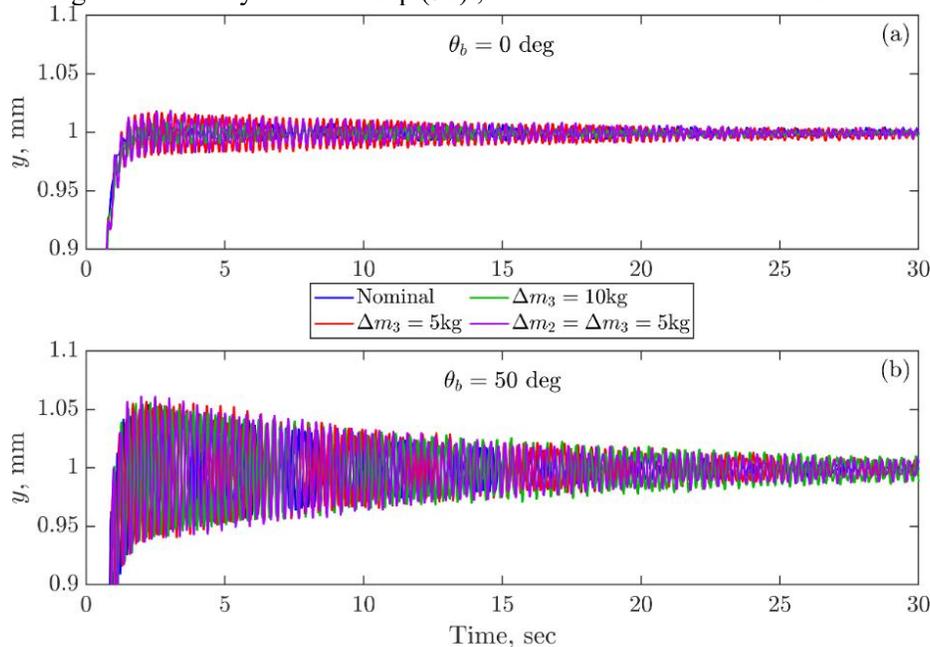

Figure 20: Step response of perturbed experimental testbed with added mass.

Table 1: Approximated first mode natural frequency,$f_1$, and damping coefficient $\zeta_1$, using the model of the perturbed physical system and estimated values of the damped vibration model (Eq. (32)) for the steady-state signals of Figure 20.

| Configuration kg | $f_1$ Hz | $\zeta_1$ % | $\theta_b$ deg | $R^2$ | $f_r$ Hz | $\zeta_r$ % | $A_{res}$ μm | $\theta_b$ deg | $R^2$ | $f_r$ Hz | $\zeta_r$ % | $A_{res}$ μm |
|---|---|---|---|---|---|---|---|---|---|---|---|---|
| Nominal | 4.40 | 0.38 | 0 | 0.86 | 4.44 | 0.19 | 6 | 50 | 0.99 | 4.44 | 0.34 | 56 |
| $\Delta m_3 = 5$ | 4.10 | 0.36 | 0 | 0.99 | 4.12 | 0.20 | 19 | 50 | 0.98 | 4.11 | 0.30 | 58 |
| $\Delta m_3 = 10$ | 3.85 | 0.33 | 0 | 0.60 | 3.85 | 0.10 | 5 | 50 | 0.99 | 3.84 | 0.28 | 56 |
| $\Delta m_2 = \Delta m_3 = 5$ | 3.98 | 0.34 | 0 | 0.89 | 3.97 | 0.33 | 15 | 50 | 0.99 | 3.97 | 0.31 | 59 |

Table 1 shows the approximated natural frequency and damping coefficient by including different lumped masses in the original model of Section 2.1. The mass matrix of the system, in this case, is given by:

$$M = \begin{bmatrix} m_1 + m_2 + m_m & 0 \\ 0 & m_3 \end{bmatrix} \quad (37)$$

The added mass values are substituted to the mass matrix if Eq. (37), then the natural frequencies and modes are obtained by solving the generalized structural eigenvalue problem:

$$(K - \omega_n^2 M)\phi_n = 0 \quad (38)$$

The modal damping coefficient is obtained as:

$$\zeta_n = \phi_n^T C \phi_n / 2\omega_n \quad (39)$$

Note that when no backlash is added, the single decaying mode model of Eq. (32) does not always fit the measured data well. This is because the control action can affect the response since the higher modes' contribution is no longer negligible at these amplitudes. Still, the single decaying mode predicted modal frequencies are almost the same for the cases with and without the backlash and are in excellent agreement with the approximated one. The estimated modal damping is more affected by the existence of the additional modes. But when the backlash is added, there is a good agreement between the approximated and estimated modal damping coefficients for all the different configurations.

Based on the experimental analysis, it is noted that the IMC-based controller can withstand significant changes to the nominal structure while keeping both the closed-loop stability and performance.

### 6.2. Mismodelling effect

To examine the effects of mismodelling, the nominal mass and stiffness parameters of the internal model, $\hat{G}_2$, identified in Section 2.2, were perturbed. The perturbed model was then used as the internal model and in the prefilter design. In the implementation of the controller, a dead zone of $\pm 0.9$ degrees was used for the motor's inner feedback.

The nominal mass was perturbed by adding $\pm 10$ and $\pm 5$ kg to it, and the nominal stiffness by adding $\pm 11$ and $\pm 5.5$ kN to it. All combinations were investigated experimentally based on a measured step response without backlash and with a virtual backlash gap angle of 50 degrees. A decaying single-mode was fitted to the steady state error based on Eq. (32). Since the physical structure is the same for all the following experiments, the identified natural frequency and damping coefficients are also the same. Therefore, only the residual vibration amplitude is reported.

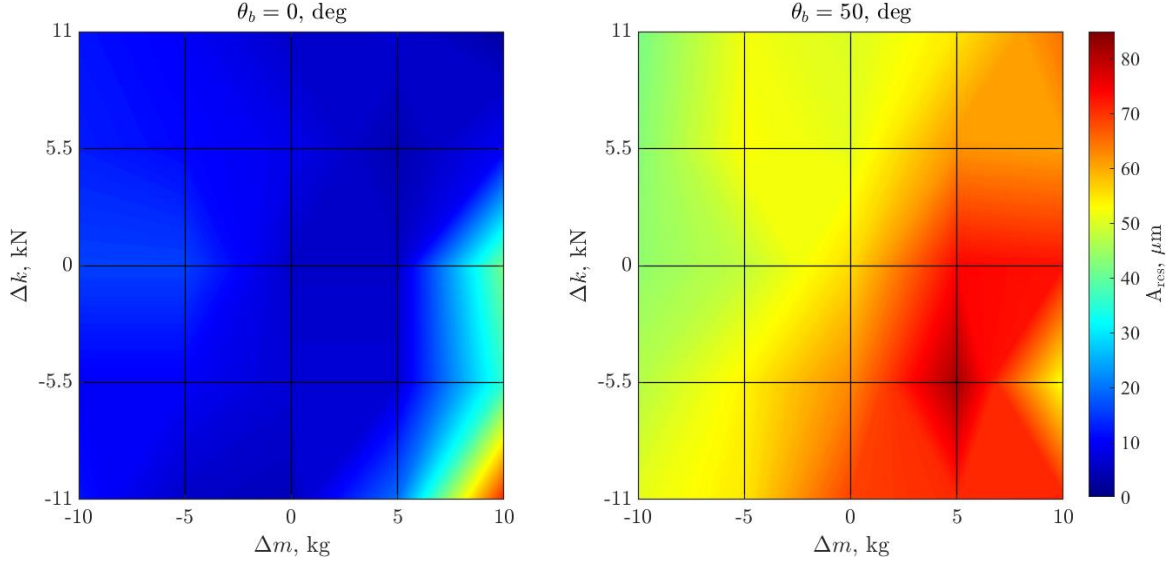
Figure 21: The effect of mismodelling on the residual vibration amplitude, $A_{res}$.

The estimated residual vibration amplitudes are shown graphically in Figure 21 for the different cases with and without backlash. Stability is maintained for all plotted cases and does not depend on the existence of the backlash, although it is lost for more extreme cases of mismatch, for example, $\Delta m = 10$kg and $\Delta k = -13$kN or $\Delta m = 15$kg and $\Delta k = -11$kN. In both combinations the internal model's natural frequency is significantly smaller than the one of the physical structure. Based on Figure 21, overestimating the mass contributes to a higher residual vibration amplitude, while underestimating the mass can improve the nominal results.

To conclude, if the initial controller design results in a stable closed-loop, the change in the amount of backlash during the system's lifetime will not affect the stability up to high levels of backlash. Only the performance is degraded due to the growing backlash. In Section 7, a possible approach to reduce the effect of the backlash on the transient response based on the designed settling time is presented. This could be combined with a novelty decision test [22] to reconfigure the controller and ensure the desired transient performance is maintained even for large backlash gap angles.

## 7. Effect of the reference model's settling time

Based on the discussion of Section 4.2, it can prove advantageous to reduce the bandwidth of the reference model in Eq. (31) for the nonlinear IMC controller. This can be done by increasing the reference model's settling time, $\tau_r$.

The transfer function of the reference model can also be written as:

$$G_r = \left(\frac{1}{\frac{\tau_r}{\tau_0}s+1}\right)^2 \qquad (40)$$

where $\tau_0 = 6.6385$ is the normalisation parameter, and $\tau_r = 1.1379$ in Eq. (31). The -3dB bandwidth frequency, in this case, can be derived as:

$$f_b = \frac{\tau_0\sqrt{\sqrt{2}-1}}{2\pi\tau_r} = \frac{4.27}{2\pi\tau_r} \text{ Hz} \qquad (41)$$

A set of experiments was conducted using the IMC-based controller with a dead zone of 0.9 degrees and when different settling times were used in the controllers' design. The response was examined for the case without backlash and for a case with a virtual backlash of 50 degrees.

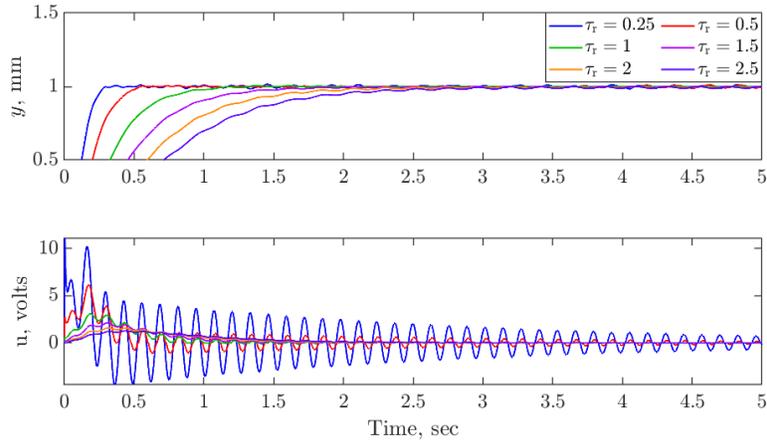
Figure 22: Step response of reference model design for different settling times with no backlash.

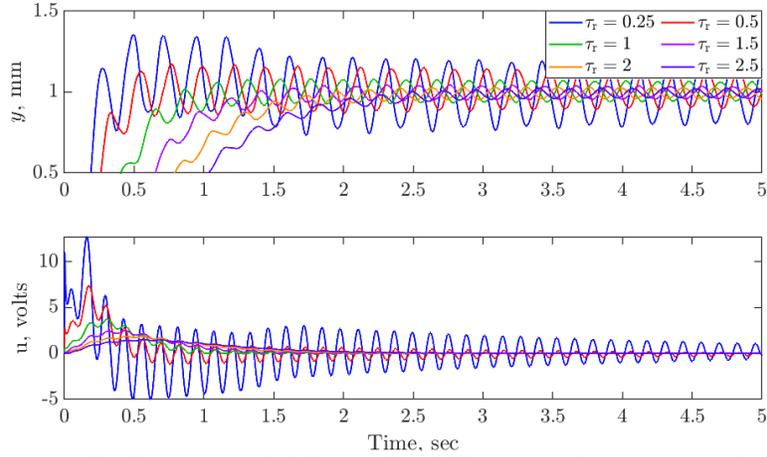
Figure 23: Step response of reference model design for different settling times with 50 degrees backlash.

Figure 22 and Figure 23 show the results for the different controllers without and with the added backlash. When there is no backlash, the response is near the steady-steady after the desired settling time, although the required control action becomes more significant for short settling times. Once the backlash is introduced, the residual vibration amplitudes are affected by the settling time. When the settling time becomes longer, the residual vibration amplitudes become smaller. The settling time could thus be chosen based on the identified amount of backlash and the allowable amount of residual vibration. For example, if the backlash gap angle is 50 degrees and the amplitude of the residual vibration is required to be less than 5% of the steady-state value, a settling time of $\tau_r = 1.5$ seconds should be used.

## 8. Conclusions

In this work, a robust model-based feedback controller is developed for a position control system with sandwiched backlash. The controller design methodology is based on the linear internal model control, IMC approach. A three-platform structure was used as the experimental testbed, for which a reduced-order model was identified experimentally to develop and test the controller. A method to introduce a controllable amount of virtual backlash between the motor and structure was developed and verified experimentally. A linear IMC-based controller was designed based on robust stability and performance criteria and compared to a model-free PID controller. The stability of the two systems was analysed analytically using the describing function approximation. Different command signals were investigated experimentally. The addition of nonlinear components to the controller was also examined. The sensitivity of the proposed controller to structural modification and mismodelling was studied experimentally. Finally, the effect of the reference model's settling time on the residual vibration amplitude was assessed.

Based on the analysis and results of Section 4, it is concluded that the IMC controller performance is significantly better than that of the model-free PID controller in the nominal case when no virtual backlash is added. This improved performance is obtained at the expense of a more sophisticated controller structure. However, unlike the PID controller, the linear IMC controller maintains stability even for high levels of backlash and retains its performance, with only small levels of residual vibration. When a smooth sinusoidal command signal is used, the output phase delay and the control amplitude are almost linear functions of the backlash gap angle, so either quantity could potentially be used to assess the amount of backlash in the system.

Based on the comparison between the linear and nonlinear IMC controllers of Section 5, it is noted that the linear IMC controller is robust and maintains stability and performance even up to high backlash angles. Moreover, it is concluded that there is no clear advantage when including a nonlinear backlash estimator with the exact model and backlash angle. On the contrary, it increases the required control power in many cases without any significant effect on the residual vibration. However, adding a small dead zone can avoid drifts inside the backlash gap and overcome the measurement noise. When the width of the dead zone is tuned appropriately, it ensures that the motor's angle remains constant in the backlash and diminishes the effect of measurement noise at the cost of a slight bias of the steady-state output.

Based on the experimental case study of Section 6, the stability of the proposed linear IMC with a small dead zone was found to be very robust to structural modification and mismodelling. To reduce the effects of backlash on performance, however, the settling time of the reference model could be adjusted at the controller design step.

To summarise, in this application, the proposed controller is highly robust to the presence of nonnegligible backlash, structural modifications, and mismodelling. The design procedure is case-dependent, however, and insights into the tuning of the different parameters were discussed. In particular, the reference model settling time and dead zone width can be adjusted to reduce the residual vibration and the effect of measurement noise at the cost of a longer settling time and steady-state bias, respectively. This illustrates the trade-offs expected in practice between the desired controller's performance, the output settling time and accuracy, and the controller's robustness to model uncertainty and noise.

Although, as stated, the design procedure is case-dependent, different methods can be developed to identify the system model and to choose the reference model used in the design based on some notion of optimality. For example, one can use a simulative optimisation to tune the settling time of the reference model that ensures some required transient performance under worst-case scenario model uncertainties. In addition, one could use learning methods to obtain the linear or the nonlinear internal model while the system is being controlled. Examples of this approach can involve utilising data-informed models, where the model is chosen using a sparse model [36] or a neural-network approach. The effect of adaptively identifying the system and consequently updating the internal model controller can be, however, challenging to analyse and predict.

## Appendix

Assuming that the backlash input is sinusoidal: $\theta_m = \Theta_m \sin\omega t$, with an amplitude $\Theta_m > \theta_b$, and defining the characteristic time $\gamma = \sin^{-1}(1 - 2\chi)$, were $\chi \equiv \theta_b/\Theta_m$ is the ratio between the backlash gap angle and the input amplitude and is bounded between zero to one. The backlash response can be divided into four zones of the normalized time $\tau = \omega t \in (\pi/2, 2\pi + \pi/2)$:

Zone I, $\pi/2 \leq \tau < \pi - \gamma$:
$$\theta_d(\tau) = \Theta_m - \theta_b = (1 - \chi)\Theta_m$$

Zone II, $\pi - \gamma \leq \tau < 3\pi/2$:
$$\theta_d(\tau) = \Theta_m \sin(\tau) + \theta_b = (\sin(\tau) + \chi)\Theta_m$$

Zone III, $3\pi/2 \leq \tau < 2\pi - \gamma$:
$$\theta_d(\tau) = -\Theta_m + \theta_b = (-1 + \chi)\Theta_m$$

Zone IV, $2\pi - \gamma \leq \tau < 2\pi + \pi/2$:
$$\theta_d(\tau) = \Theta_m \sin(\tau) - \theta_b = (\sin(\tau) - \chi)\Theta_m$$

Introducing the Fourier harmonic expansion

$$\theta_d(\tau) = \frac{a_0}{2} + \sum_{n=1}^{\infty} a_n \cos n\tau + b_n \sin n\tau \quad \text{(A.1)}$$

$$a_n = \frac{1}{\pi}\int_0^{2\pi} \cos n\tau\, \theta_d(\tau) d\tau \quad \text{(A.2)}$$

$$b_n = \frac{1}{\pi}\int_0^{2\pi} \sin n\tau\, \theta_d d\tau \quad \text{(A.3)}$$

It follows after integration that:

$$a_1 = \frac{4\theta_b}{\pi}\left(\frac{\theta_b}{A} - 1\right) = \frac{4\theta_b}{\pi}(-1 + \chi) \quad \text{(A.4)}$$

and

$$b_1 = \frac{A}{\pi}\left(\frac{\pi}{2} + \gamma + 2\left(1 - 2\frac{\theta_b}{A}\right)\sqrt{\frac{A - \theta_b}{A^2}\theta_b}\right) = \frac{A}{\pi}\left(\frac{\pi}{2} + \gamma + 2(1 - 2\chi)\sqrt{\chi(1 - \chi)}\right) \quad \text{(A.5)}$$

The complex describing function is thus given by:

$$N(\chi) = \frac{1}{A}(b_1 + i a_1) = \frac{1}{\pi}\left(\frac{\pi}{2} + \gamma + 2(1 - 2\chi)\sqrt{\chi(1 - \chi)}\right) + i\frac{4}{\pi}\chi(-1 + \chi) \quad \text{(A.6)}$$

The describing function amplitude and phase are shown in Figure A.1.

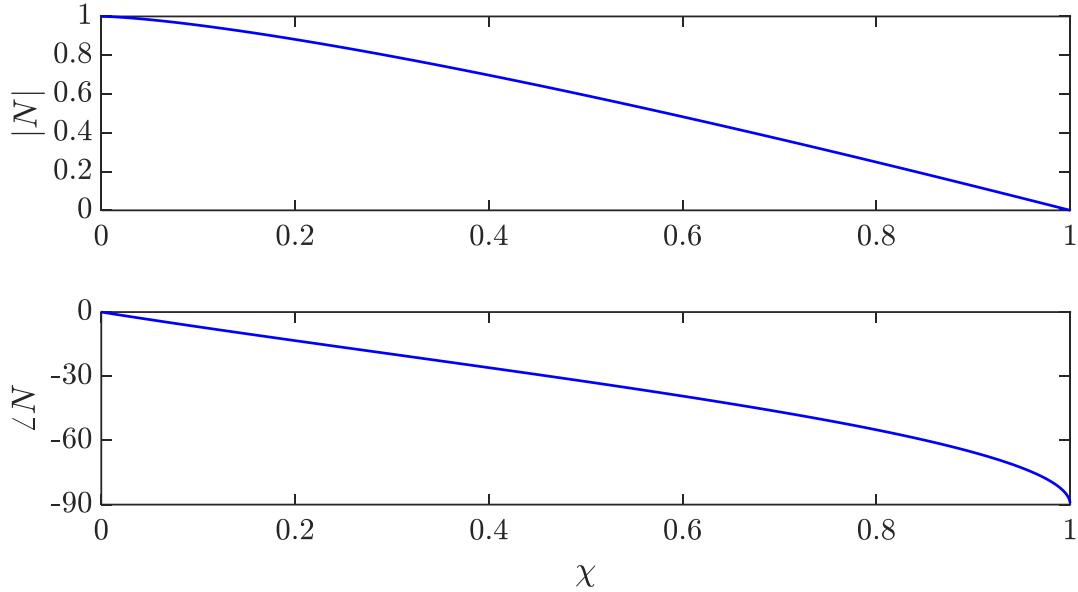

Figure A. 1: Backlash characteristic function, magnitude and phase, versus the ratio of backlash gap angle to the backlash input angle amplitude.

# Acknowledgement

The authors gratefully acknowledge the support of the UK Engineering and Physical Sciences Research Council (EPSRC) through the DigiTwin project (grant EP/R006768/1).

# References

[1] Tao G, Kokotovic P. Adaptive control of systems with actuator and sensor nonlinearities. John Wiley & Sons, Inc.; 1996.
[2] Boneh R, Yaniv O. Reduction of limit cycle amplitude in the presence of backlash. Journal of Dynamic Systems, Measurement and Control, Transactions of the ASME 1999;121:278–84. https://doi.org/10.1115/1.2802466.
[3] Lagerberg A. A literature survey on control of automotive powertrains with backlash. 2001.
[4] Nordin M, Gutman PO. Controlling mechanical systems with backlash—a survey. Automatica 2002;38:1633–49. https://doi.org/10.1016/S0005-1098(02)00047-X.
[5] Dong R, Tan Y, He D. Nonsmooth identification of mechanical systems with backlash-like hysteresis. Journal of Control Theory and Applications 2013 11:3 2013;11:477–82. https://doi.org/10.1007/S11768-013-2106-Y.
[6] Alban LE. Systematic analysis of gear failures. ASM International; 1985.
[7] Vered Y, Elliott SJ. The use of digital twins to remotely update feedback controllers for the motion control of nonlinear dynamic systems. Mech Syst Signal Process 2023;185:109770. https://doi.org/https://doi.org/10.1016/j.ymssp.2022.109770.
[8] Tao G, Ma X, Ling Y. Optimal and nonlinear decoupling control of systems with sandwiched backlash. Automatica 2001;37:165–76. https://doi.org/10.1016/S0005-1098(00)00153-9.
[9] Seidl DR, Lam SL, Putman JA, Lorenz RD. Neural Network Compensation of Gear Backlash Hysteresis in Position-Controlled Mechanisms. IEEE Trans Ind Appl 1995;31:1475–83. https://doi.org/10.1109/28.475744.
[10] Rostalski P, Besselmann T, Barić M, Van Belzen F, Morari M. A hybrid approach to modelling, control and state estimation of mechanical systems with backlash. Int J Control 2007;80:1729–40. https://doi.org/10.1080/00207170701493985.
[11] Kalman RE. A new approach to linear filtering and prediction problems. Journal of Fluids Engineering, Transactions of the ASME 1960;82:35–45. https://doi.org/10.1115/1.3662552.
[12] Luenberger DG. Observers for Multivariable Systems. IEEE Trans Automat Contr 1966;11:190–7. https://doi.org/10.1109/TAC.1966.1098323.
[13] Sariyildiz E, Oboe R, Ohnishi K. Disturbance Observer-Based Robust Control and Its Applications: 35th Anniversary Overview. IEEE Transactions on Industrial Electronics 2020;67:2042–53. https://doi.org/10.1109/TIE.2019.2903752.
[14] Morari M, Zafiriou E. Robust process control. Prentice Hall; 1989.
[15] Khalil I, Doyle J, Glover K. Robust and optimal control. 1996.
[16] Morari M, Zafiriou E. Robust process control. Prentice Hall; 1989.
[17] Elliott SJ. Signal Processing for Active Control. Elsevier; 2001. https://doi.org/10.1016/b978-0-12-237085-4.x5000-5.
[18] Saidi L, Dibi Z, Ababsa F. Internal model control and disturbance observers. WSEAS Transactions on Circuits and Systems 2006;5:525–9. https://doi.org/10.1109/CDC.1991.261294.
[19] Zames G. Feedback and Optimal Sensitivity: Model Reference Transformations, Multiplicative Seminorms, and Approximate Inverses. IEEE Trans Automat Contr 1981;26:301–20. https://doi.org/10.1109/TAC.1981.1102603.

[20]   Garcia CE, Morari M. Internal Model Control. 1. a Unifying Review and Some New Results. Industrial and Engineering Chemistry Process Design and Development 1982;21:308–23. https://doi.org/10.1021/I200017A016.

[21]   Morari M. Internal Model Control - Theory and Applications. IFAC Proceedings Volumes 1983;16:1–18. https://doi.org/10.1016/S1474-6670(17)64183-1.

[22]   Vered Y, Elliott SJ. The use of a digital twin to reconfigure a nonlinear internal model controller. ISMA2022-USD2022 Proceedings, Leuven, Belgium: 2022.

[23]   Bucher I, Mirkin L, Vered Y. Input Shaping via FIR L2 Preview Tracking. Proceedings of the IEEE Conference on Decision and Control, vol. 2020- Decem, 2020. https://doi.org/10.1109/CDC42340.2020.9304337.

[24]   Goubej M, Vyhlídal T, Schlegel M. Frequency weighted H2 optimization of multi-mode input shaper. Automatica 2020;121. https://doi.org/10.1016/j.automatica.2020.109202.

[25]   Youla DC, Bongiorno JJ, Jabr HA. Modern Wiener-Hopf Design of Optimal Controllers — Part II: The Multivariable Case. IEEE Trans Automat Contr 1976;21:319–38. https://doi.org/10.1109/TAC.1976.1101223.

[26]   Vidyasagar M. Robust stabilization of singularly perturbed systems. Syst Control Lett 1985;5:413–8. https://doi.org/10.1016/0167-6911(85)90066-0.

[27]   Economou CQ, Morari M, Palsson BO. Internal model control: extension to nonlinear system. Industrial and Engineering Chemistry Process Design and Development 1986;25:403–11. https://doi.org/10.1021/I200033A010.

[28]   Qiuping H, Rangaiah GP. Adaptive internal model control of nonlinear processes. Chem Eng Sci 1999;54:1205–20. https://doi.org/10.1016/S0009-2509(98)00543-0.

[29]   Landau ID, Lozano R, M'Saad M, Karimi A. Adaptive control: algorithms, analysis and applications. Springer Science & Business Media; 2011.

[30]   Datta A. Adaptive internal model control. Springer Science & Business Media; 2012.

[31]   Dong R, Tan Y, He D. A nonsmooth IMC method for mechanical systems with backlash. J Control Theory Appl 2013;11:600–7. https://doi.org/10.1007/s11768-013-2113-z.

[32]   Dong R, Tan Y, He D. A nonsmooth IMC method for mechanical systems with backlash. J Control Theory Appl 2013;11:600–7. https://doi.org/10.1007/s11768-013-2113-z.

[33]   Tao G, Kokotovic P V. Adaptive control of systems with backlash. Automatica 1993;29:323–35. https://doi.org/10.1016/0005-1098(93)90126-E.

[34]   MathWorks. PID Tuner 2022. https://uk.mathworks.com/help/control/ref/pidtuner-app.html.

[35]   Slotine J, Li W. Applied nonlinear control. Prentice hall Englewood Cliffs, NJ; 1991.

[36]   Brunton SL, Proctor JL, Kutz JN. Sparse Identification of Nonlinear Dynamics with Control (SINDYc). IFAC 2016;49:710–5. https://doi.org/10.1016/j.ifacol.2016.10.249.